# On-chip visible light communication-band metasurface modulators with niobium plasmonic nano-antenna arrays

Kaveh Delfanazari [1,*] & Otto L. Muskens [2]

1. James Watt School of Engineering, University of Glasgow, Glasgow G12 8QQ, UK

2. Physics and Astronomy, University of Southampton, Southampton SO17 1BJ, UK

*corresponding author: kaveh.delfanazari@glasgow.ac.uk

**We introduce chip-integrated visible light communication-band modulators based on niobium (Nb) metallic plasmonic nano-antenna arrays. Our plasmonic nano-devices provide strong sensitivity to the polarization of the incident visible light and the geometrical parameters of their subwavelength nanoscale building blocks. Moreover, they offer optical modulation properties with modulation depth MD $\cong$ 60% at resonant wavelength $\lambda$= 716 nm, at room temperature. By engineering the photo response of the Nb nano-device arrays, we observe a maximum extinction $A(\lambda)$= 1- $R(\lambda)$ $\cong$ 95 % at resonant wavelength $\lambda$= 650 nm. Our results suggest that the integrated Nb nano-antenna array devices can be considered as suitable platforms for the realisation of chip-scale optoelectronic devices interfacing cryogenics quantum circuits, and fibre-based communication systems, for applications in quantum computing, quantum communication, and quantum processing.**

Niobium (Nb), a metallic superconductor with quantum mechanics phases below their superconducting transition temperature $T_c \cong 9$ K, has been widely used in superconducting quantum technology. Superconducting quantum circuits have shown potential as a leading technology for quantum computing [1-10]. Cryogenic photonic links using optical fibres with their low thermal conductivity have been recently introduced to surpass the heat load associated with the growing number of quantum processors in low temperatures and to connect superconducting quantum hardware nodes with fibre-based telecommunication platforms [6-8]. Nevertheless, the optical properties of superconducting circuits, superconducting nanostructures, and the devices interfacing superconductors and photonics are surprisingly unexplored [6-10]. Here, at the first step, the photo-response of niobium (Nb) nano-antenna arrays of different geometries is studied both numerically (using COMSOL Multiphysics) and experimentally (using a microspectrophotometer CRAIC QDI2010, and an ultrafast laser with pulse duration 400 fs) at room temperature.

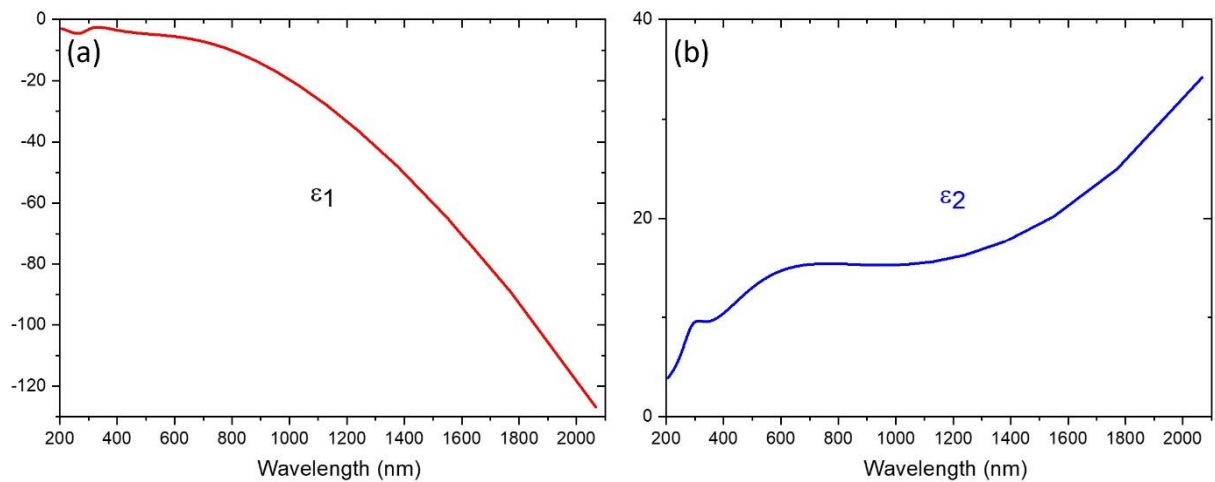

**Figure 1:** the dielectric function of the unstructured Nb film at room temperature, measured by variable-angle spectroscopic ellipsometry in the visible and optical light communication frequencies. Plasmonic property of Nb is shown for a broad frequency range.

To fabricate the integrated nano-devices, Nb thin film of 300 nm thickness was sputtered onto a sapphire substrate of 30 mm diameter and 0.5 mm thickness. To explore the optical properties of unstructured Nb thin film and to be able to simulate the photoresponse of our nano-devices in COMSOL, the variable-angle spectroscopic ellipsometry measurement was performed in the wavelengths ranges from λ= 200 nm to 2000 nm at room temperature and down to liquid nitrogen temperature (not shown here) using a Horiba Uvisel-2 spectroscopic ellipsometer [9]. Figure 1 shows the Nb plasmonic properties and a negative permittivity which was observed from UV to NIR frequency ranges. The superconducting transition temperature $T_c$ of our unstructured Nb thin films was confirmed to be around $T_c \cong 9$ K through the DC transport measurements (not shown here). The integrated nano-antenna arrays were patterned by focused ion beam milling FIB (FEI Helios NanoLab 600). In total, we fabricated five chips, each with an array of Nb subwavelength elements of different geometries, each array covering a total area of 100×100 µm². The light-matter interaction experiments were done when the Nb chip was loaded in closed-cycle optical cryostats with temperature tuning range capability between $T$= 293 K and 3 K.

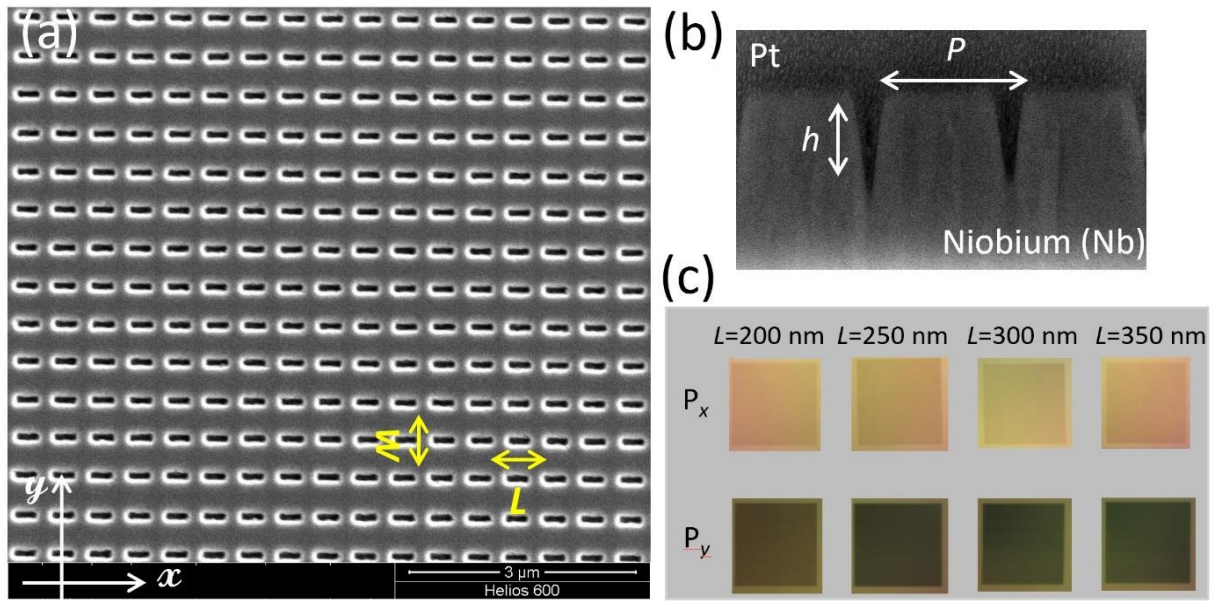

**Figure 2:** (a) Scanning Electron Microscope (SEM) image of the Nb integrated nanocircuitry with superconducting transition temperature $T_c \cong 9$ K. (b) SEM of the cross section of the part of the Nb nano-antenna array showing the antenna profile with a period $P \cong 500$ nm and a depth $h \cong 200$ nm. Platinum (Pt) layer was deposited, after optical spectroscopy, for protection during sectioning for device characterisation purposes. (c) optical microscope image of four chips with arrays of plasmonic Nb nano-antennas at two different polarization angles. Plasmonic colours for the optical nano-antenna arrays depends on the meta-atom geometrical parameters, and the angle of the incident light.

We first present the data for four chips that contain an array of nano-antennas of rectangle (slit) geometries. In the chips, the slit lengths vary between $L \cong$ 200 nm, 250 nm, 300 nm, and 350 nm for a fixed width $w \cong$ 50 nm. Figure 2 (a) illustrates an example of the integrated nano-device fabricated by using FIB. Figure 2 (b) shows the SEM image of the cross-section of part of the array. The optical nano-antenna array has a period of $P \cong$ 500 nm and a depth of $h \cong$ 200 nm. The platinum (Pt) layer was deposited, after optical spectroscopy, for protection during sectioning for device characterization purposes. The nano-antenna arrays provide plasmonic colors when illuminated with white light. Figure 2 (c) shows the optical microscope images of four chips with collections of plasmonic Nb nano-antenna arrays at incident light polarizations parallel ($P_x$) and perpendicular ($P_y$) to the slits. Plasmonic colors for the optical nano-antenna arrays change with each meta-atom length and with the incident light polarization.

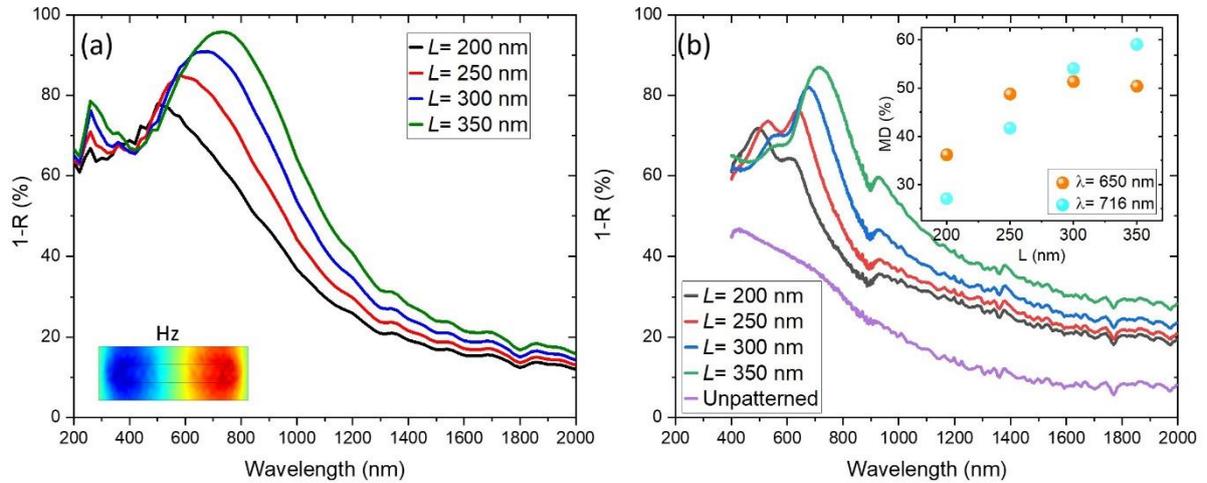

**Figure 3:** (a) The numerical simulation, and (b) experimental resonant response of various nano-antenna arrays with length varying from $L=$ 200 nm to 350 nm for light polarized perpendicular to the slits ($P_y$). Inset to Fig. 3 (a) is the $H_z$-field distribution in the nano-antenna at the resonance wavelength $\lambda=$ 650 nm.

The optical Nb nano-devices were numerically modeled using a fully three-dimensional finite element technique (COMSOL Multiphysics). The simulation results for four chips are shown in Figure 3 (a) for y-polarized plane waves ($P_y$). The $H_z$-field distribution in the nano-antenna at the resonance wavelength $\lambda=$ 650 nm is shown in the inset. The resonance shows an electrical dipole response in the field distribution in the groove. Figure 3 (b) shows the experimental results for nano-devices that were measured in a microspectrophotometer under light polarized perpendicular to the slits ($P_y$). We observe extinction spectra A(λ)= 1- R(λ) at visible light communication bands, and the resonances are found to move to longer wavelengths (redshift) with increasing nano-antenna length. Relatively good agreement can be seen between the experiment and simulation, taking into account the *V*-shaped profile cross-section of the nano-antennas made by FIB. A clear modulation of visible light extinction spectra could therefore be observed in Nb nano-devices by engineering the meta-atom geomatical parameters.

The modulation depth (MD) is defined as [11,12]

$$MD = \frac{A_{Nb\ nanoantenna\ array} - A_{unpatterened\ Nb\ film}}{A_{Nb\ nanoantenna\ array}}$$

where $A_{unpatterened\ Nb\ film}$ and $A_{Nb\ nanoantenna\ array}$ are maximum extinction of the unpatterned Nb thin film, and Nb nano-antenna arrays with length varying from *L*= 200 nm to 350 nm, respectively. We find an apparent increase of MD with increasing the length of nano-antenna arrays. MD for *L*= 350 nm, reaches the maximum value of 60 % at the resonant wavelengths λ= 716 nm.

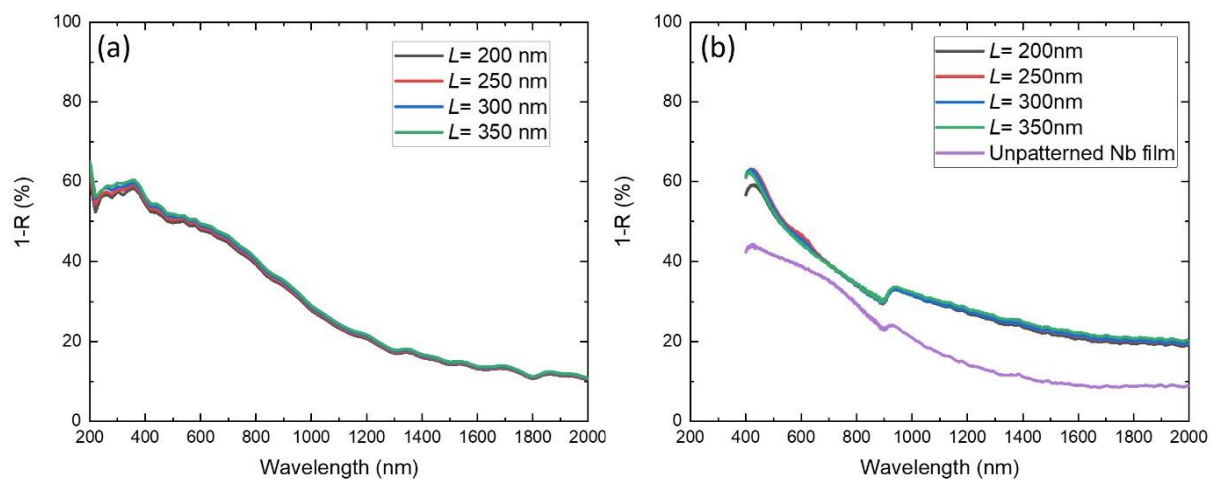

**Figure 4:** (a) The numerical simulation, and (b) experimental resonant response of various nanocircuit arrays with length varying from *L*= 200 nm to 350 nm for light polarized parallel to the slits, ($P_x$).

We find that the extinction spectra A(λ)= 1- R(λ) are pretty different between unstructured (un-patterned) Nb thin film and nano-antenna arrays for incident light polarization perpendicular to the slits ($P_y$). We did not observe any resonance peaks from un-patterned Nb thin film ( see Fig. 3 (b)), indicating the importance of nanostructure arrays engraved into the surface of Nb thin film in modulation and enhancement of light-matter interactions. The relatively narrow plasmon resonances that are modulated by nano-antenna arrays length of $L=200$ nm to 350 nm, for light polarized perpendicular to the slits, radically change the optical response of Nb thin film and their corresponding plasmonic colors in the visible range without a requirement to chemical modification or usage of a secondary material coating, as shown in Fig. 2 (c).

The numerical simulation and experimental measurement of the Nb nano-antenna array for light polarized parallel to the slits ($P_x$) are presented in Fig. 4. We find a good agreement between simulation and experimental results. The A(λ) spectra for Nb nano-antenna arrays show a strong sensitivity to the polarization of incident light, while the photoresponse is quite small for unstructured Nb thin film. This again confirms the vital role of plasmonic nanostructures in engineering the light-matter interactions in photonic nano-circuitry.

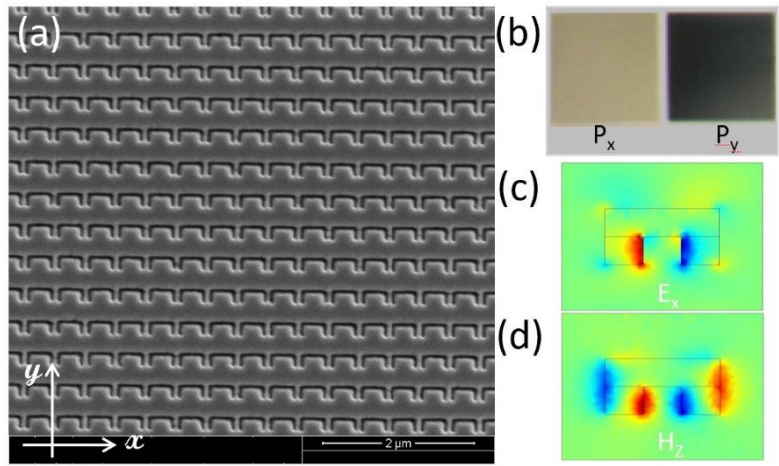

**Figure 5:** (a) SEM image of the Nb integrated plasmonic nano-antennas with SRR geometries. (b) optical microscope image of the arrays of plasmonic Nb nano-antennas at two different polarization, parallel ($P_x$) and perpendicular ($P_y$) to the device plane. Plasmonic colors in structured Nb films show a strong sensitivity to the geometry of the nano-antenna element and to the polarization of the incident light. Nearfield profiles of electric (c) and magnetic (d) modes at wavelength λ= 650 nm.

To further investigate the effect of geometrical parameters on visible light modulation properties of Nb plasmonic nanostructures, the fifth chip was patterned with a nano-antenna array whose cavity shape is selected to be a U-Shape split-ring resonator (SRR). The chip was milled by FIB into the Nb film surface, and the high fabrication quality and uniformity of the samples were confirmed by SEM images as shown in Fig. 5 (a). Plasmonic colors of the arrays are shown in Fig. 5 (b) for two different polarizations. Similar to nano-slit antenna arrays, they show a dramatic change in color compared to that of the unstructured Nb metal thin films. The chip was loaded into an optical cryostat and measured with a microspectrophotometer at room temperature. The experimental results for light polarized parallel ($P_x$) and perpendicular ($P_y$) to the split ring resonator are presented in Fig. 6. A huge difference was found between the extinction spectra $A(\lambda) = 1 - R(\lambda)$ of unstructured Nb thin film (Fig. 6 (a)) and the nano-split ring resonator arrays (Fig. 6 (b)). The resonance response of the plasmonic Nb nano-split ring resonators is clearly a strong function of the geometry and size of the meta-atoms.

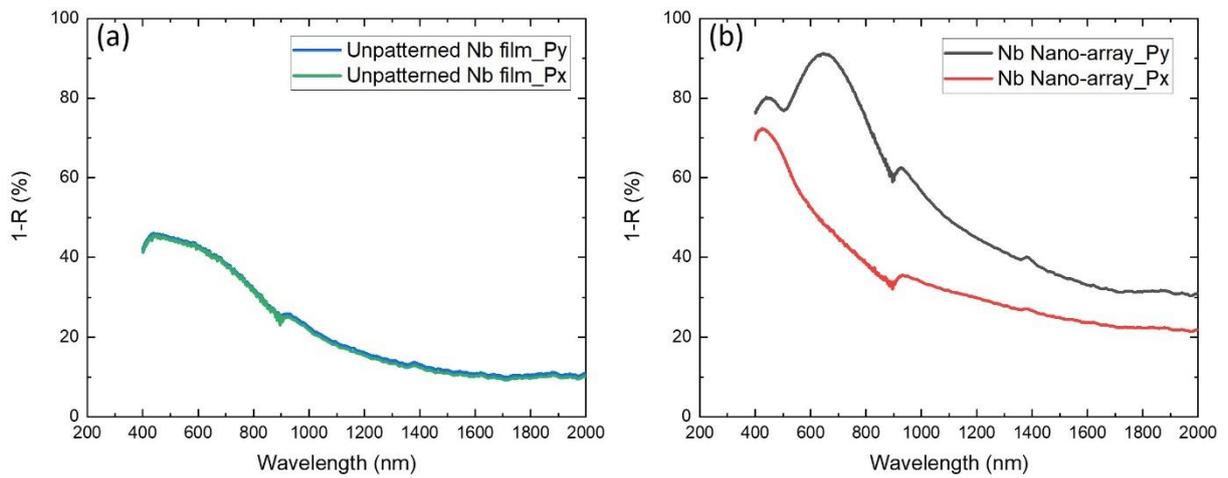

**Figure 6:** (a) experimental resonant response of (a) unstructured Nb thin film, and (b) nano-antenna split ring resonator array for light polarized perpendicular ($P_y$) and parallel ($P_x$) to the device plane.

Nb nano-split ring resonators contain both electric and magnetic modes. The simulated nearfield profiles of magnetic and electric modes of the nano-SRR with a U-shaped hole structure milled through the Nb metallic thin film are shown in Fig. 5 (c) and Fig. 5 (d), respectively. The parallel dipoles in the figure indicating plasmon mode. There is no induced dipole that can be seen along the top arm of the SRR, and the $H_z$ field is only localized around the side arms because the same charge polarity is stimulated on the Nb nano-antenna sidearms [13,14]. Robust localization of $E_y$-field in the Nb nano-antenna element indicates that plasmon mode is dominated by vertical dipoles. This suggests that plasmon mode in the Nb nano-antenna split ring resonator array can only be excited when the incident light is $P_y$ polarized [13-15].

We also find that the nano-split ring resonator array chip offers a very sensitive photoresponse to the incident light polarization. We observe nearly 80% of visible light absorptance for a wide wavelength ranges between $\lambda$= 400 nm and 800 nm, which corresponds to visible light communication band, while this value reduces by more than 40 % for unstructured Nb thin film in the same wavelengths, as shown in Fig. 6. The nano-device shows a resonance peak at $\lambda$= 650 nm, with > 90 % light absorptance, make it an excellent ingredient for integration with Nb-based superconducting quantum hardware, and visible light communication system elements, such as polymer optical fibre (POF) with their robustness under bending and stretching. Nb photonic nano-devices integrated with POF-based, or similar technology readout platforms with low thermal conductivity [6-8], may also help to reduce the heat load associated with the growing number of quantum hardware in cryogenic temperatures. Such plasmonic nano-devices in different geometrical parameters can also be integrated with other quantum devices, such as quantum light sources, hybrid light modulators, and slow light devices to improve their optical functionalities [16-35]. Our approach may also help the development of on-chip integrated nano-photonic circuits and devices [36-60].

**Conclusion:**

We fabricated plasmonic nano-antenna arrays of different geometries on the facet of Nb thin films by using FIB milling. Our Nano-device arrays offer optical modulation of light at visible communication bands with a maximum modulation depth MD≅ 60% at resonant wavelength λ= 716 nm, at room temperature. We observe maximum extinction spectra A(λ)= 1- R(λ) ≅ 95 % at wavelength λ= 650 nm by engineering the Nb nano-device arrays geometrical parameters. Metallic nanostructures based on Nb become superconductors with a quantum mechanic phase around *T*= 9 K, offering potential thermo-photo modulation of visible light at cryogenic temperatures. Such nano-devices have the potential to be integrated with Nb-based superconducting quantum hardware and fiber-based telecommunication read-out systems, paving the way for the realisation of chip-scale all superconducting photonic links, modulators, and filters for applications in quantum computing and communication.